# SERpredict: Detection of tissue- or tumor-specific isoforms generated through exonization of transposable elements

Britta Mersch[1], Noa Sela[2], Gil Ast[2], Sándor Suhai[1] and Agnes Hotz-Wagenblatt*[1]

Address: [1]Department of Molecular Biophysics, German Cancer Research Center (DKFZ), Heidelberg, Germany and [2]Department of Human Molecular Genetics, Sackler Faculty of Medicine, Tel Aviv University, Tel Aviv, Israel

Email: Britta Mersch - b.mersch@dkfz.de; Noa Sela - noasela@post.tau.ac.il; Gil Ast - gilast@post.tau.ac.il; Sándor Suhai - s.suhai@dkfz.de; Agnes Hotz-Wagenblatt* - hotz-wagenblatt@dkfz.de

* Corresponding author

## Abstract

**Background:** Transposed elements (TEs) are known to affect transcriptomes, because either new exons are generated from intronic transposed elements (this is called *exonization*), or the element inserts into the exon, leading to a new transcript. Several examples in the literature show that isoforms generated by an exonization are specific to a certain tissue (for example the heart muscle) or inflict a disease. Thus, exonizations can have negative effects for the transcriptome of an organism.

**Results:** As we aimed at detecting other tissue- or tumor-specific isoforms in human and mouse genomes which were generated through exonization of a transposed element, we designed the automated analysis pipeline *SERpredict* (SER = <u>S</u>pecific <u>E</u>xonized <u>R</u>etroelement) making use of Bayesian Statistics. With this pipeline, we found several genes in which a transposed element formed a tissue- or tumor-specific isoform.

**Conclusion:** Our results show that *SERpredict* produces relevant results, demonstrating the importance of transposed elements in shaping both the human and the mouse transcriptomes. The effect of transposed elements on the human transcriptome is several times higher than the effect on the mouse transcriptome, due to the contribution of the primate-specific Alu elements.

## Background

Transposed elements (TEs) are sequences of DNA that can move from one position to another in the genome. There are two classes of transposed elements, the DNA transposons and the retroelements. DNA transposons usually move by cut and paste using the transposase enzyme. In contrast, retroelements are genetic elements that integrate in a genome via an RNA intermediate which is reverse-transcribed to DNA. In mammals, almost half the genome is comprised of TEs: around 45% of the human genome is made up of them. This translates to millions of elements, so that on average, every gene in our genome contains about 3 transposed elements. Transposed elements comprise approximately 37% of the mouse genome.

The human and mouse genome sequences show that TEs have played an important role in shaping the genomes [1,2]. The human genome contains retroelements such as Alu, which is a short interspersed element (SINE), MIR (mammalian interspersed repeat) as well as LINE-1 (L1), LINE-2 (L2) and CR1 (L3). The last three of the given families of retroelements are termed long interspersed elements. In addition, the human genome contains LTR elements such as MaLR (mammalian apparent LTR-retrotransposon), ERVL and ERV1 (endogenous retroviruses) as well as DNA transposons where common families are MER1 and MER2. The mouse genome contains MIR elements as well as rodent-specific SINEs such as B1 (homologous to the left arm of the Alu), B2, B4 and ID as well as LINEs such as L1, L2 and CR1. Similar to the human genome, the mouse genome contains LTRs and DNA transposons. With approximately 1 million copies, Alu is the most frequently encountered TE in the human genome. In mouse, B1 and L1 are the elements with the highest number of copies (B1: 500,000 copies, L1: 800,000 copies).

Through splicing processes ("exonizations"), small pieces of transposed elements can be inserted into mature mRNAs. These exonizations are caused by motifs that resemble consensus splice sites in both strands of the TEs [3]. The transposed elements do not only contain these splice sites but also polyadenylation sites, promoters, enhancers and silencers. Therefore, they can add a variety of functions to their targeted genes [4-6].

Mutations within intronic TEs may yield active splice sites which can be used instead of the normal splice sites, leading to the partial exonization of the intronic TE. However, the other TEs of the human and mouse genomes can be exonized, too. In a previous study, Sela et al. [7] showed that 1824 TEs are exonized in the human genome, of which about 58% are Alus. In the mouse genome, 506 transposed elements are exonized, most of which are either B1 or L1 elements (26% and 20%, respectively). Thus, transposed elements can affect the transcriptome. Either new exons are generated from intronic TEs (see Figure 1a (i)), or the TE inserts into the first or last exon of a gene (Figure 1b (i)), leading to a new transcript [8]. In the first case, the exonization can either generate an internal cassette exon (Figure 1a (ii)), an alternative 3'splice site (Figure 1a (iii)), an alternative 5'splice site (Figure 1a (iv)) or a constitutively spliced exon (Figure 1a (v)) [7,9]. In the case of insertions into first or last exons, the insertions cause either an elongation of the first/last exon (Figure 1b (ii, iii)) or an activation of an alternative intron (Figure 1b (iv)). For the exact number of occurrences of the different events please refer to [7].

It has been previously reported that more than 5% of the alternatively spliced exons in the human genome are Alu-derived and that all Alu-derived exons are the result of exonization of intronic sequences [8]. It can therefore be supposed that genetic diseases can occur when an intronic TE is constitutively or alternatively spliced into the mature mRNA. Searching the literature indeed uncovers evidence that Alu insertions cause genetic disorders [10-13].

Another effect of new exonizations is a potential tissue specificity, in which an exon shows strong tissue regulation [14]. An experimental verification of this mechanism is described in a report on Alu de-novo insertion and subsequent exonization within the dystrophin gene that creates a tissue-specific exon inflicting cardiomyopathy [15].

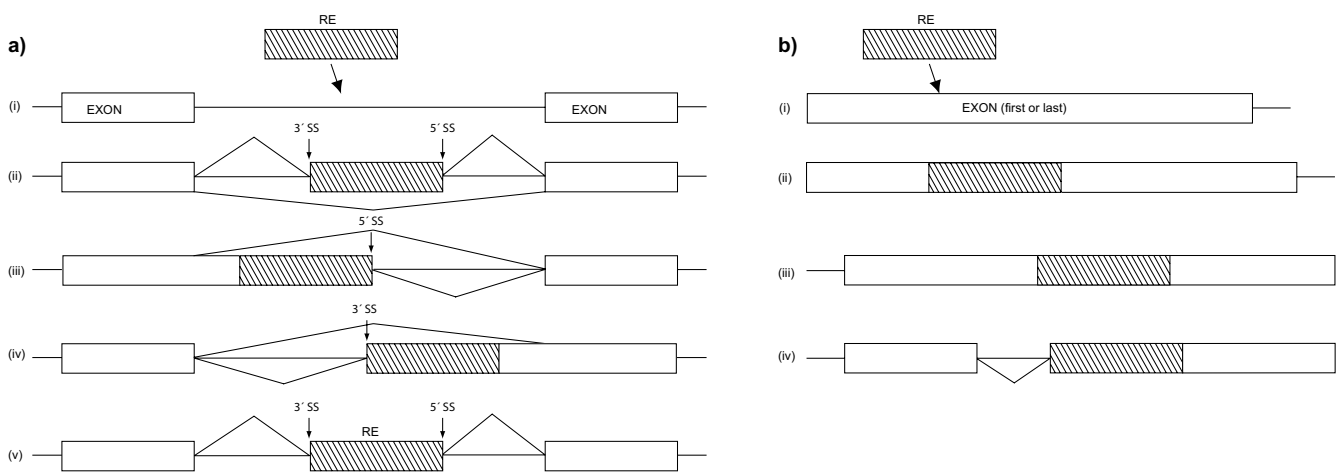

**Figure 1**
**The effects of TE insertions. a)** (i) TE inserts into an intron of a gene. (ii-v) show the possible effects of this integration; (ii) alternatively exon is created, (iii) TE contributes alternative 5'splice site, (iv) TE contributes alternative 3'splice site, (v) TE creates a constitutively spliced exon. **b)** (i) TE inserts into the first or last exon of a gene. (ii – iv) show the possible effect of this integration: (ii, iii) enlargement of first or last exon, (iv) TE activates an alternative intron.

Furthermore, since tumorous tissues have been shown to adopt aberrant splicing patterns [16], there might be TE exonizations that are potentially tumor-specific. The survivin gene is one example in which an Alu-generated splice variant is tumor-specific [17].

For the detection of new tissue- or tumor-specific TE-containing isoforms in human or mouse genes, we designed and implemented *SERpredict*, an analysis pipeline making use of Bayesian Statistics.

## Implementation

*SERpredict* is based on several databases: the Ensembl database [18], the UCSC genome browser database [19], dbEST [20], and EMBL [21]. How they are used and combined is described in the following section.

### Library classification

To obtain information about the tissue and the health status of alternative splice forms of genes, the databases dbEST [20] for expressed sequence tags (ESTs) and EMBL [21] for mRNAs are used. These EST and mRNA sequence databases provide information about tissue and tumor sources. For dbEST, library information which include an ID, the library name, the organism, the tissue and a more detailed description is provided with each entry. For the EMBL database, there are features termed "clone lib" and "tissue type". However, this poses a problem since the names of the tissues and tumors are not standardized across the databases. For this reason, we extracted all the EST and mRNA identifiers from the two databases dbEST and EMBL, obtained the associated library information and assigned a tissue category to every given tissue according to [22] (see as well Additional file 1). Here, we used key words to identify 52 different cell or tissue source categories, e.g., leukocyte is mapped to the category blood, hippocampus is mapped to brain and so on. Furthermore, either "tumor" or "normal" was added to each library, using again keywords. All the information was then stored in a locally installed MySQL database which is automatically updated if one of the underlying databases is updated. The "annotated" EST and mRNA sequences obtained in this way were used to perform the statistical analysis to determine whether a certain isoforms is tissue- or tumor-specific.

### Tissue and tumor specificity

The analysis for tissue or tumor specificity of a certain alternative splice form can be done using the above "annotated" EST and mRNA sequences. Therefore, all EST and mRNA sequences which map the gene of interest have to be extracted. To determine tissue or tumor specificity, the extracted sequences have to be divided into two groups reflecting the two isoforms of the gene. This is easy in our special case of alternative isoforms because one of the isoforms was generated by the exonization of an TE. On the basis of this information, the ESTs and mRNAs are separated into the groups "holding", when the TE is present in the sequence of interest or "skipping", in cases where the TE is not. Using the library classification terms for the sequences, we then get four sets of distributions. For each of those EST and mRNA sequences skipping the TE as well as for those holding the TE, we obtain a tissue and a source (tumor or normal) distribution.

Determining tissue or tumor specificity from these distributions is not easy, because tissue and tumor source data for EST or mRNA sequences are often incomplete and inconsistent. For a certain gene there are often only a few ESTs sequenced from a particular tissue covering the exons of interest. We therefore have to cope with a poor EST library coverage. Additionally, there are extremely different numbers of ESTs and mRNAs for the different tissues, see Figure 2. This leads to a sampling bias problem. To address these problems, statistical analysis is needed.

Furthermore, when dealing with tissue or tumor specificity there is a problem with including ESTs from cell lines into the analysis. Cell lines are often immortalized, and the immortal lines obtained might not be a perfect representation of the original cells in primary culture. For an estimation how many of the ESTs originated from cell lines, we checked the annotation in the dbEST database and determined that only about 10% of the human and mouse EST sequences were derived from cell lines.

### Statistical analysis

To deal with the incomplete and inconsistent data, we used a previously described Bayesian statistics approach to identify tissue-specific exons [23] and to identify exons showing deregulated splicing in tumors [24].

To identify tissue-specific exons, a tissue specificity score (TS score) is computed. The confidence that a certain splice variant is preferred in tissue T is calculated as a Bayesian posterior probability:

$$P(\theta_{1T} > 50\% \mid obs) = \frac{\int_{0.5}^{1} P(obs|\theta_{1T})P(\theta_{1T})d\theta_{1T}}{\int_{0}^{1} P(obs|\theta_{1T})P(\theta_{1T})d\theta_{1T}}$$

(1)

Here, $\theta_{1T}$ represent the hidden frequency of a splice variant in a specific tissue T and *obs* stands for the number of ESTs and mRNAs observed in tissue T. $P(obs|\theta_{1T})$ is calculated using a binomial distribution and $P(\theta_{1T}) = 1$ was used as uninformative prior probability. In the same way, the posterior probability that the same splice variant is preferred in the pool of all other tissues ~ is computed. The TS score for tissue T is then defined as the difference between the posterior probability for tissue T and the posterior probability for the pool of all other tissues:

$$TS = 100\ [P(\theta_{1T} > 50\%|obs) - P(\theta_{1\sim} > 50\%|obs)] \quad (2)$$

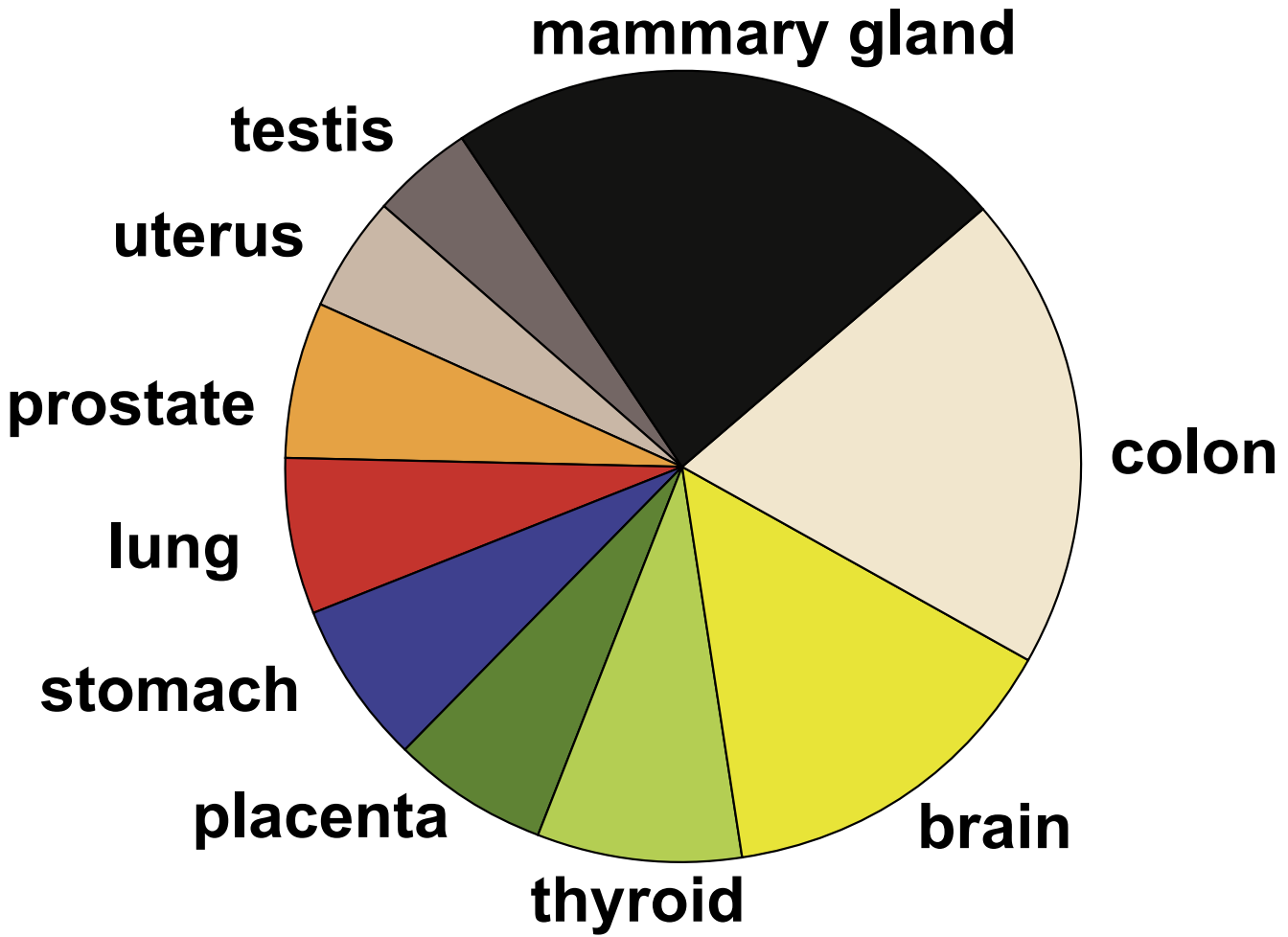

**Figure 2**
**Pie chart of EST numbers**. Number of ESTs for the different tissues. This figure indicates the extremely different numbers of ESTs and mRNAs for the different tissues.

Here, $\theta_{1\sim}$ is the frequency of a splice variant in the pool of all other tissues ~. To assess the stability of the TS score robustness values, $r_{TS}$ and $r_{TS\sim}$ were calculated analogous to the "jack-knife" resampling method. For more details, please refer to [23].

To identify tumor-specific exons, a log-odd score (LOD score) is calculated, giving the confidence that the frequency of a splice variant in tumor tissue ($\theta_T$) is higher than the frequency of the splice variant in normal tissue ($\theta_N$):

$$LOD = -\log_{10} \frac{P(\theta_T > \theta_N)}{1 - P(\theta_T > \theta_N)} \quad (3)$$

The LOD score was calculated using direct numerical integration [24].

The criteria used for high-confidence of tissue specificity were TS > 50, $r_{TS}$ > 0.9 and $r_{TS\sim}$ > 0.9 as described in [23]. A necessary condition for high confidence tissue specificity was at least three EST observations of the mRNA containing the exon in tissue T. As we wanted to have results with high significance only, we changed the criteria for the TS score to TS > 85 for our pipeline. For tumor specificity, a log-odd score was calculated. As in [24], a log-odd score above 2, equivalent to a p-value <0.01, was considered significant.

### Work flow in SERpredict

Using the information presented in the sections above, we designed *SERpredict* to detect tissue- or tumor-specific isoforms, which were generated through the exonization of transposed elements. The work flow is displayed in Figure 3.

Initially, the genomic information of the input sequence is determined. Therefore, a Blast search [25] with Ensembl_cdna [18] is performed. Utilizing the Ensembl Application Programming Interface (API), the extracted Ensembl gene identifier is then used to find all transcripts of the gene and thereby all the different exons. If there is no Blast hit matching the criteria (Identity > 95% and E-value <$10^{-3}$), an empty output is produced.

Subsequently, every extracted exon is screened for transposed elements. This is done using the chrN_rmsk table of the UCSC genome browser database [19], which maps the positions of all TEs that have been found by RepeatMasker [26] and the Repbase annotations [27,28] to the human and mouse genomes, respectively. This approach is much faster than using RepeatMasker directly.

Finally, for every such TE-containing exon, an analysis of tissue or tumor specificity is performed as described in the Section "Statistical Analysis". Subsequently, *SERpredict* extracts all expressed sequence tags (ESTs) and mRNA sequences from the UCSC genome browser database. These are then divided into two groups as described in the Section "Tissue and tumor specificity" and used as input for the statistical analysis.

As output, *SERpredict* returns a file with the following information:

• Information about the genomic location: Ensembl gene identifier, gene description, chromosome, strand, start and end on genome, number of transcripts and corresponding number of exons

• A graphical display of the alternative splice forms of the gene

• Information about the repetitive elements: family, ID of exon in which the TE is located, start and end of the TE and the IDs of the transcript containing the TE-exon

• If observed: the tissue or tumor specificity of the TE-containing exon

These results are provided as HTML for visual inspection (see Figure 4) or can be downloaded as XML for easier extraction of relevant results and for storage in private databases.

### Results and Discussion

All annotated genes of the human and mouse genomes were screened for TE-containing exons. The number of times the different transposed elements were exonized (and fulfilled the condition of at least three EST observations of the mRNA containing the exon in tissue T) are shown in Table 1 (for the human genome) and Table 2 (for the mouse genome).

The 859 human and 260 mouse TE-containing exons were then analyzed for tissue or tumor specificity using *SERpredict*. In the human exon list, we were able to identify 39 tissue-specifically spliced exons (see Table 3 for the exons with tissue specificity (TS) score > 90). In the mouse exon list, 11 exons showed tissue-specific splicing (see Table 4 for the exons with tissue specificity (TS) score > 90). In the human genome, 18 exons belonged to Alu, 5 were L1 exons, 2 were L2 exons, 1 was an CR1 exon, 5 were MIR exons, 4 were LTR exons and 4 were exons derived from DNA transposons. The highest amount of tissue-specific exonizations arises from the exonization of an Alu element. The fact that the Alu is the most abundant transposed element in the human genome and that it contains potential splice sites makes it a much better-suited sequence for the exonization process than other transposed elements [7] and could be a reason for these results. In mouse, 4 were B1 exons, 2 were B2 exons and 2 were LTR exons. For B4, L2, MIR there was one exon each. The higher amount of specific exonized B1 elements is consistent with the fact that B1 derived from the same ancestral origin as Alu. Still, B1 does not reach the same amount of specific exonizations as Alu because the majority of exonizations of Alu occur in the right arm of the Alu element which is not present in B1. In contrast to the dimeric structure of the Alu element, B1 is a monomer.

We did not observe a tendency for specificity in any certain tissue in humans. In the mouse genome, interestingly, there is a bias for specific exons in pancreas tissue. This is not due to a bias in the number of ESTs/mRNAs from mouse pancreatic tissue in the database since there are as many pancreatic sequences as sequences from other tissues like intestine or blood. Therefore, this is an interesting result for which we do not have any explanations so far.

As MIR SINEs were active prior to the mammalian diversification [29] it was unexpected to find 5 tissue-specific MIR exonizations in human and only 1 in the mouse genome. We examined the orthologous loci of the 5 relevant genes RDH13, Elmo2, MRRF, Tri14 and NP_060401.2 in the human and the mouse genome and discovered that there is no MIR element in the mouse genome in 4 of the 5 cases. Only for MRRF there is a MIR in the mouse genome but the exonization in mouse is not

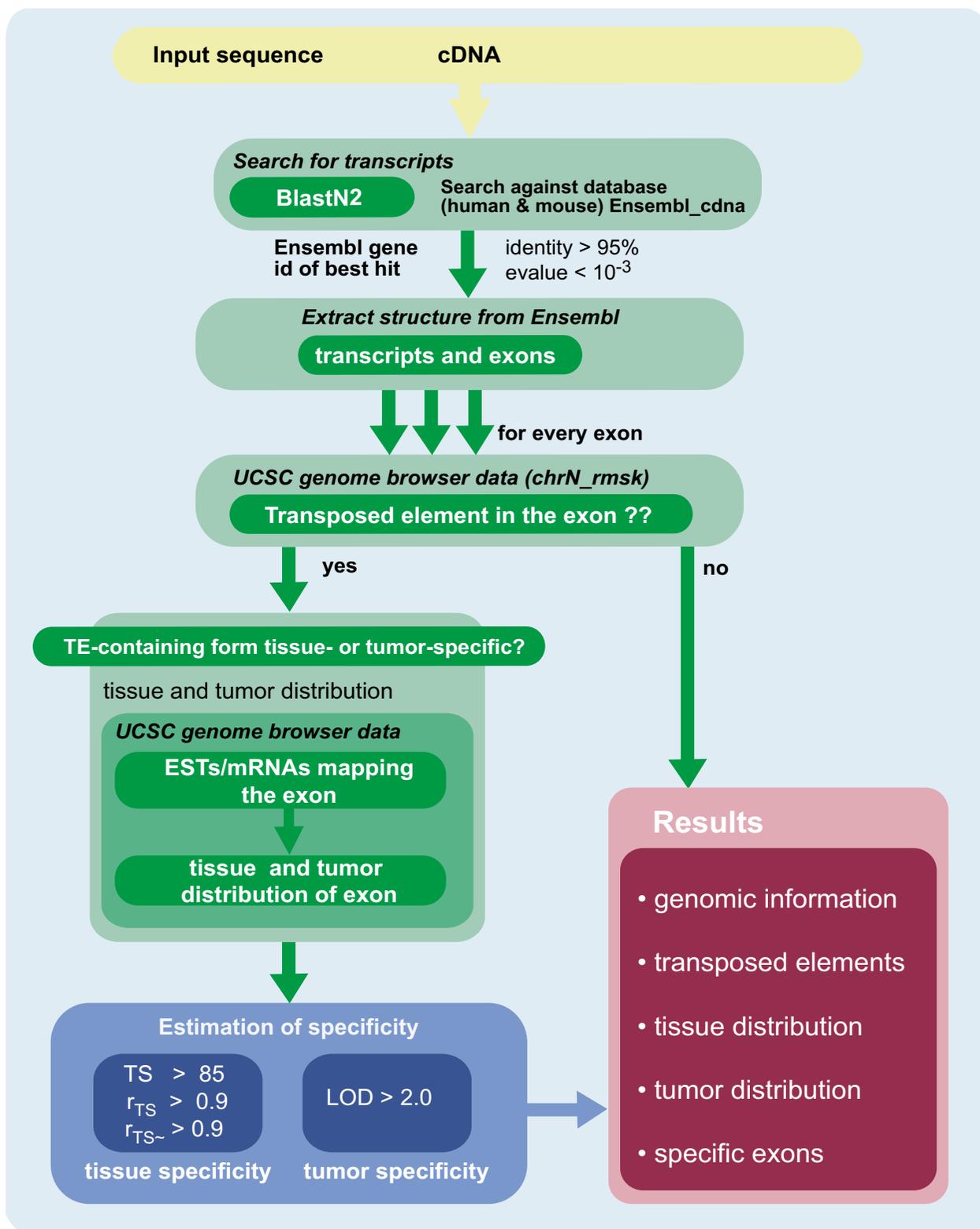

**Figure 3**
**Work flow of SERpredict**. Programs and rules used for extracting tissue- or tumor-specific TE-containing exons, for details see Section "Work flow in SERpredict".

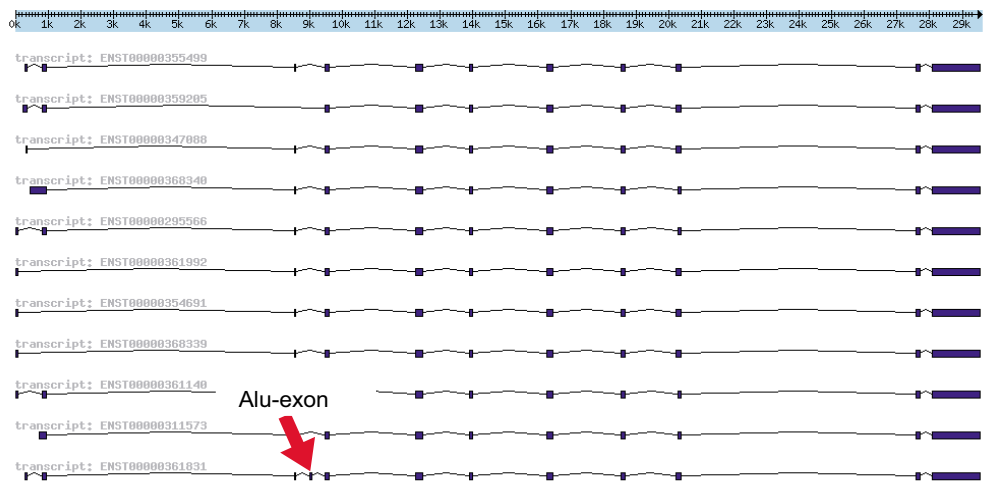

**Figure 4**
**Example output of SERpredict**. The output of *SERpredict* for [EMBL:AF466401] is presented. One of the isoforms shows a tumor-specific exon which was generated through the exonization of an Alu element.

**Table 1: Exonizations in the human genome**

| Human | | | | | | | |
|---|---|---|---|---|---|---|---|
| Transposed element | Alu | L1 | L2 | CR1 | MIR | LTR | DNA |
| Number of exonizations | 432 | 88 | 40 | 4 | 86 | 119 | 90 |

Number of exonized transposed elements in the human genome which have a least three EST observations of the mRNA containing the TE exon. The Alu element is exonized most frequently among TEs.

tissue-specific. For the specific exon of gene ST6galrsc4 in the mouse genome, there is a MIR at the same position in the human genome but the exon boundaries are different. Therefore, the MIR is not exonized in the human genome.

To show the efficiency of *SERpredict*, some of the genes which we predicted to have a tissue-specific TE-derived exon were verified by searching both the literature and the database annotations. Isoform 2 of the T-cell activation NFKB-like protein contains an Alu exon and was predicted as ovary-specific (Table 3), which was verified through the human SwissProt [30] entry Q9BRG9. A testis-specific isoform of TPK1 (Thiamine pyrophosphokinase 1) is described in the OMIM [31] entry 606370 [32,33]. This isoform is 100 bp longer than the broadly expressed variant. This complies with our results of an additional ERV1-derived exon of about 100 bp which makes this isoform testis-specific (Table 3). Additionally, the 4F2 cell-surface heavy chain protein seems to be highly expressed in the early stage of new bone formation [34]. Although we found an alternative isoform expressed in bone (Table 3), the specificity of the TE-derived exon is not described in the literature and could therefore not be verified.

Our second analysis identified exons which were spliced in a tumor-specific way. We found 21 such exons in human and 2 in mouse genes. In the human genome, 11 were Alu exons, 1 was a L1 exon, 1 was a L2 exon and 4 were MIR exons, 3 were LTR exons and one exon derived from a DNA transposon (see Table 5). In mouse, there was 1 L1 exon and 1 MIR exon (see Table 6). The data was filtered to search for exons that were intronic within normal tissues and were recognized as exons only within tumorous tissues and, as such, could serve as potential markers for tumor diagnostics. One such exon which contains an Alu element was found in the human gene YY1AP1 (YY1-associated protein 1: hepatocellular carcinoma susceptibility protein). All results for TS > 85 and LOD > 2 are given in Additional file 2 for the human and the mouse genome.

We also found an indication for the accuracy of our predictions of genes with tumor-specific exons. From the ST6GALNAC6 gene a 2.4 kB transcript has been described for colon carcinoma, while in normal colon transcripts of 2.5 and 7.5 kB length are found [35,36]. The colon carcinoma transcript could represent the isoform which omits the first exon and contains the tumor-specific exon.

Taking these results into account, *SERpredict* is a useful tool for analyzing TE insertions in genes and to determine their effects for the human and mouse transcriptomes. On the one hand, their insertion into mature mRNAs and the subsequent change in the protein can cause effects in single tissues or even cause major illnesses like cancer. This has already been shown in several examples in the literature [10-12,15,17]. On the other hand, these new exons could be raw material for future evolution of the organisms. The new alternative TE-exons are only included into a fraction of the transcripts of a gene while the rest of the transcripts maintain their original function. Therefore, the addition may be free to evolve with no loss of original function. If the alternative form gains a useful function, its splice sites are strengthened or it can become tissue-specific if the new function has only local benefits [14].

**Table 2: Exonizations in the mouse genome**

| Mouse | | | | | | | | | |
|---|---|---|---|---|---|---|---|---|---|
| transposed element | B1 | B2 | B4 | L1 | L2 | ID | MIR | LTR | DNA |
| number of exonizations | 55 | 33 | 22 | 37 | 9 | 4 | 20 | 72 | 8 |

Number of exonized transposed elements in the mouse genome which have a least three EST observations of the mRNA containing the TE exon. The B1 element, which is homologous to the left arm of the Alu element, is exonized most frequently among mouse TEs.

**Table 3: Human tissue-specific TE-exons**

| Gene | TE | Chr. | Tissue | TS score |
|---|---|---|---|---|
| Zinc finger protein 195 | L1 | 11 | nerve | 92.41 |
| Zinc finger protein 33A | L1 | 10 | trachea | 96.81 |
| Ribonuclease P protein subunit p38 | Alu | 10 | bone | 92.30 |
| AP-3 complex subunit mu-1 | Alu | 10 | eye | 99.21 |
| Zinc finger protein 195. | Alu | 11 | nerve | 92.41 |
| 4F2 cell-surface antigen heavy chain | Alu | 11 | bone | 96.69 |
|  |  |  | uterus | 96.34 |
| Suppressor of G2 allele of SKP1 homolog | Alu | 13 | pancreas | 94.39 |
|  |  |  | muscle | 96.65 |
| Centrosomal protein of 27 kDa | Alu | 15 | uterus | 93.29 |
| Fumarylacetoacetate hydrolase domain-containing protein 1 | Alu | 16 | placenta | 94.01 |
| T-cell activation NFKB-like protein | Alu | 19 | ovary | 93.65 |
| Zinc finger protein 320 | Alu | 19 | embryo | 96.68 |
| Zinc finger MYM-type protein 1 | Alu | 1 | brain | 97.84 |
| Protein MANBAL | Alu | 20 | stomach | 93.75 |
| Serine/threonine-protein kinase 6 | Alu | 20 | mouth_oral | 94.16 |
| CDNA FLJ20699 fis. clone KAIA2372. | Alu | 22 | muscle | 93.75 |
|  |  |  | brain | 98.69 |
| DNA directed RNA polymerase II | Alu | 7 | colon | 93.45 |
|  |  |  | ovary | 93.45 |
|  |  |  | thymus | 98.29 |
| Putative ribosomal RNA methyltransferase 2 | Alu | 7 | breast | 93.74 |
| GPI ethanolamine phosphate transf. 3 | Alu | 9 | eye | 91.82 |
| Soluble calcium-activated nucl. 1 | L2 | 17 | placenta | 98.14 |
| Intraflagellar transport 20 homolog | L2 | 17 | colon | 98.92 |
|  |  |  | muscle | 93.73 |
| CDNA FLJ32655 fis | CR1 | 17 | testis | 99.99 |
| Retinol dehydrogenase 13 | MIR | 19 | testis | 93.74 |
|  |  |  | uterus | 92.69 |
| Tripartite motif-containing protein 14. | MIR | 9 | thymus | 93.39 |
| Salivary alpha-amylase precursor | ERVK | 1 | muscle | 97.67 |
| Thiamin pyrophosphokinase 1 | ERV1 | 7 | testis | 93.75 |
| Hypothetical protein Q8WZ27 | ERV1 | 4 | thyroid | 93.73 |
| Trafficking protein particle complex protein 2 | MER1 | X | testis | 92.55 |
| mTERF domain-containing protein 2. | MER2 | 2 | brain | 99.09 |

Potentially tissue-specific TE-exons in the human transcriptome. From left to right: the gene name in which the exonization occurred, the transposed elements family name, the chromosome number, the name of the tissue to which the exon is specific and the TS score.

**Table 4: Mouse tissue-specific TE-exons**

| Gene | TE | Chr. | Tissue | TS score |
|---|---|---|---|---|
| RIKEN cDNA 9830124H08 gene | B1 | 14 | pancreas | 99.21 |
| Gametogenetin binding protein 1 | B1 | 17 | pancreas | 90.57 |
| G protein-coupled receptor 177 | B1 | 3 | pancreas | 98.05 |
| Hydroxysteroid dehydrogenase like 2 | B1 | 4 | pancreas | 98.44 |
| NFKB inhibitor interacting Ras-like protein 1 | B2 | 14 | pancreas | 96.48 |
| RIKEN cDNA 4930444A02 gene | B2 | 8 | pancreas | 92.81 |
| ADP-ribosylation factor-like 4A | B4 | 12 | pancreas | 93.75 |
| Cyclin-dependent kinase 2 | L2 | 10 | pancreas | 92.75 |
| Target of EGR1. member 1 | L2 | 4 | limb | 92.75 |
| ST6 | MIR | 2 | limb | 92.74 |
|  |  |  | intestine | 99.4 |

Potentially tissue-specific TE-exons in the mouse transcriptome. From left to right: the gene name in which the exonization occurred, the transposed elements family name, the chromosome number, the name of the tissue to which the exon is specific and the TS score.

**Table 5: Human tumor-specific TE-exons**

| Gene | TE | Chr. | LOD |
|---|---|---|---|
| Centrosomal protein of 27 kDa | Alu | 15 | 2.93 |
| G-protein coupled receptor 56 precursor | Alu | 16 | 2.8 |
| T-cell activation NFKB-like protein | Alu | 19 | 2.76 |
| Dipeptidyl peptidase 9 | Alu | 19 | 3.49 |
| NADH dehydrogenase | Alu | 19 | 2.07 |
| YY1-associated protein 1 | Alu | 1 | 3.15 |
| Centromere protein R | Alu | 1 | 2.74 |
| Selenoprotein T precursor. | Alu | 3 | 2.64 |
| Putative ribosomal RNA methyltransferase 2 | Alu | 7 | 2.4 |
| GPI ethanolamine phosphate transferase 3 | Alu | 9 | 2.07 |
| Protein RM11 homolog. | Alu | 9 | 2.03 |
| NHP2-like protein 1 | L1 | 22 | 9.1 |
| Zinc finger protein DZIP1 | L2 | 13 | 2.34 |
| Dynein light chain 2A. cytoplasmic | MIR | 20 | 4.43 |
| Protein NipSnap1. | MIR | 22 | 6.55 |
| ST6 | MIR | 9 | 2.69 |
| Tripartite motif-containing protein 14. | MIR | 9 | 2.86 |
| 40S ribosomal protein SA | ERV1 | 3 | 12.27 |
| DNA directed RNA polymerase II | MaLR | 7 | 2.34 |
| CDNA FLJ33708 fis, clone BRAWH2007862. | MaLR | 6 | 2.18 |
| Beta-microseminoprotein precursor | MER1 | 10 | 2.92 |

Potentially tumor-specific TE-exons in the human transcriptome. From left to right: the gene name in which the exonization occurred, the transposed elements family name, the chromosome number and the LOD score.

## Conclusion

Our results show that *SERpredict* produces relevant results, demonstrating the importance of transposed elements in shaping both the human and the mouse transcriptomes. Due to the contribution of the primate-specific Alu elements, the effect of TEs on the human transcriptome is several times higher than the effect on the mouse transcriptome. We found some evidences for our results in both the literature and the database annotations. Other results still need biological verification. The pipeline can therefore be used as an indicator for biologists interested in tissue- or tumor-specific isoforms to decide which gene might be interesting for further research.

Due to the incompleteness of the present gene databases, our analysis remains confined to the annotated gene data. With the continuous updating of the mRNA and EST databases, and with it our internal MySQL database, the analysis can be repeated. This will make analyses more precise and will provide results on previously undiscovered exons, using *SERpredict* to obtain either tissue- or tumor-specific splicing.

In further studies we will include additional organisms into *SERpredict* in order to determine differences to the human and mouse genomes. Additionally, we are planning to build a database containing the data of TE-containing exons, the annotation with the TEs, as well as tissue and tumor specificities for different organisms. This will be an extension and an update of the AluGene database [37,38].

## Availability and requirements

Project name: *SERpredict*

**Table 6: Mouse tumor-specific TE-exons**

| Gene | TE | Chr. | LOD |
|---|---|---|---|
| Methylmalonic aciduria | L1 | 8 | 2.76 |
| ST6 | MIR | 2 | 2.56 |

Potentially tumor-specific TE-exons in the mouse transcriptome. From left to right: the gene name in which the exonization occurred, the transposed elements family name, the chromosome number and the LOD score.

Project home page: http://genius.embnet.dkfz-heidelberg.de/menu/biounit/open-husar/

Operating system: Platform independent

Programming language: Perl; Other requirements: Browser

License: NA

Any restrictions to use by non-academics: None

### Usage

As part of the HUSAR open server, applications are listed on the web page with additional information about the tasks they perform. Query sequences can be uploaded by the usual "copy & paste" procedure into the input box. If more than one sequence is to be queried, a multiple FASTA file can be used. The query starts by clicking on the "submit" button and then the "run" button on the following page. Results can be received by selecting the tab "Go to results page". For further explanation, a flow chart, an example output, and a test sequence are given on the web page.

### Input/output formats

*SERpredict* accepts only nucleotide sequences as input. For output, see Section "Work flow in SERpredict".

### Performance

Calculations are normally fast, depending on the length of the input sequence and the number of exons the input sequence contains. A calculation takes approximately one minute.

## Abbreviations

TE – transposed element, SINE – short interspersed element, LINE – long interspersed element, MIR – mammalian interspersed repeat, EST – expressed sequence tag, TS score – tissue specificity score, LOD score – log-odd score, HUSAR – Heidelberg Unix Sequence Analysis Resources

## Authors' contributions

BM designed and programmed the pipeline, made the tests and drafted the manuscript. NS participated in designing the pipeline and provided test sequences. GA conceived of the study. SS provided guidance and helped to finish the manuscript. AH supervised the whole project, participated in designing the pipeline and helped to draft the manuscript. All authors read and approved the final manuscript.

## Additional material

> **Additional File 1**
> *Keywords for tissue categories.* Excel file with information about the keywords used to retrieve cell and tissue source information to assign a tissue category (first column) to every given tissue in the databases dbEST and EMBL.
> Click here for file
> [http://www.biomedcentral.com/content/supplementary/1471-2156-8-78-S1.xls]
>
> **Additional File 2**
> *All tissue and tumor specificities.* Excel file giving all tissue and tumor-specific TE-containing exons from the human and the mouse genome. There are 4 sheets, one for human tissue-specific TE-containing exons, one for human tumor-specific TE-containing exons, one for mouse tissue-specific TE-containing exons and the last for mouse tumor-specific TE-containing exons.
> Click here for file
> [http://www.biomedcentral.com/content/supplementary/1471-2156-8-78-S2.xls]

## Acknowledgements
This work was supported in part by the Cooperation Program in Tumor Research of the German Cancer Research Center (DKFZ) and Israeli's Ministry of Science and Technology (MOST) under grant Ca 119.

notransferase: a role for Alu elements in human mutation.** *Proc Natl Acad Sci USA* 1991, **88(3):**815-819.
12. Vervoort R, Gitzelmann R, Lissens W, Liebaers I: **A mutation (IVS8+0.6kbdelTC) creating a new donor splice site activates a cryptic exon in an Alu-element in intron 8 of the human beta-glucuronidase gene.** *Hum Genet* 1998, **103(6):**686-693.
13. Wallace M, Andersen L, Saulino A, Gregory P, Glover T, Collins F: **A de novo Alu insertion results in neurofibromatosis type 1.** *Nature* 1991, **353:**864-866.
14. Modrek B, Lee CJ: **Alternative splicing in the human, mouse and rat genomes is associated with an increased frequency of exon creation and/or loss.** *Nat Genet* 2003, **34(2):**177-180.
15. Ferlini A, Galié N, Merlini L, Sewry C, Branzi A, Muntoni F: **A novel Alu-like element rearranged in the dystrophin gene causes a splicing mutation in a family with X-linked dilated cardiomyopathy.** *Am J Hum Genet* 1998, **63(2):**436-446.
16. Wang Z, Lo HS, Yang H, Gere S, Hu Y, Buetow KH, Lee MP: **Computational analysis and experimental validation of tumor-associated alternative RNA splicing in human cancer.** *Cancer Res* 2003, **63(3):**655-657.
17. Mola G, Vela E, Fernández-Figueras MT, Isamat M, Muñoz-Mármol AM: **Exonization of Alu-generated splice variants in the survivin gene of human and non-human primates.** *J Mol Biol* 2007, **366(4):**1055-1063.
18. Hubbard TJP, Aken BL, Beal K, Ballester B, Caccamo M, Chen Y, Clarke L, Coates G, Cunningham F, Cutts T: **Ensembl 2007.** *Nucleic Acids Res* 2007:D610-D617.
19. Hinrichs AS, Karolchik D, Baertsch R, Barber GP, Bejerano G, Clawson H, Diekhans M, Furey TS, Harte RA: **The UCSC Genome Browser Database: update 2006.** *Nucleic Acids Res* 2006:D590-D598.
20. Boguski MS, Lowe TM, Tolstoshev CM: **dbEST-database for "expressed sequence tags".** *Nat Genet* 1993, **4(4):**332-333.
21. Kulikova T, Akhtar R, Aldebert P, Althorpe N, Andersson M, Baldwin A, Bates K, Bhattacharyya S, Bower L, Browne P, Castro M: **EMBL Nucleotide Sequence Database in 2006.** *Nucleic Acids Res* 2007:D16-D20.
22. Pan Q, Bakowski MA, Morris Q, Zhang W, Frey BJ, Hughes TR, Blencowe BJ: **Alternative splicing of conserved exons is frequently species-specific in human and mouse.** *Trends Genet* 2005, **21(2):**73-77.
23. Xu Q, Modrek B, Lee C: **Genome-wide detection of tissue-specific alternative splicing in the human transcriptome.** *Nucleic Acids Res* 2002, **30(17):**3754-3766.
24. Xu Q, Lee C: **Discovery of novel splice forms and functional analysis of cancer-specific alternative splicing in human expressed sequences.** *Nucleic Acids Res* 2003, **31:**5635-5643.
25. Altschul SF, Gish W, Miller W, Myers EW, Lipman DJ: **Basic local alignment search tool.** *J Mol Biol* 1990, **215(3):**403-410.
26. **RepeatMasker Home Page**   [http://www.repeatmasker.org]
27. Jurka J: **Repbase Update: a database and an electronic journal of repetitive elements.** *Trends Genet* 2000, **16(9):**418-420.
28. Jurka J, Kapitonov VV, Pavlicek A, Klonowski P, Kohany O, Walichiewicz J: **Repbase Update, a database of eukaryotic repetitive elements.** *Cytogenet Genome Res* 2005, **110(1–4):**462-467.
29. Smit AF, Riggs AD: **MIRs are classic, tRNA-derived SINEs that amplified before the mammalian radiation.** *Nucleic Acids Res* 1995, **23:**98-102.
30. Boeckmann B, Bairoch A, Apweiler R, Blatter MC, Estreicher A, Gasteiger E, Martin MJ, Michoud K, O'Donovan C, Phan I, Pilbout S, Schneider M: **The SWISS-PROT protein knowledgebase and its supplement TrEMBL in 2003.** *Nucleic Acids Res* 2003, **31:**365-370.
31. **Online Mendelian Inheritance in Man, OMIM**   [http://www.ncbi.nlm.nih.gov/omim/]
32. Nosaka K, Onozuka M, Kakazu N, Hibi S, Nishimura H, Nishino H, Abe T: **Isolation and characterization of a human thiamine pyrophosphokinase cDNA.** *Biochim Biophys Acta* 2001, **1517:**293-297.
33. Zhao F, Gao F, Goldman ID: **Molecular cloning of human thiamin pyrophosphokinase.** *Biochim Biophys Acta* 2001, **1517(2):**320-322.
34. Kim SG, Ahn YC, Yoon JH, Kim HK, Park SS, Ahn SG, Endou H, Kanai Y, Park JC, Kim DK: **Expression of amino acid transporter LAT1 and 4F2hc in the healing process after the implantation of a tooth ash and plaster of Paris mixture.** *In Vivo* 2006, **20(5):**591-597.
35. Okajima T, Chen HH, Ito H, Kiso M, Tai T, Furukawa K, Urano T, Furukawa K: **Molecular cloning and expression of mouse GD1alpha/GT1aalpha/GQ1balpha synthase (ST6GalNAc VI) gene.** *J Biol Chem* 2000, **275(10):**6717-6723.
36. Tsuchida A, Okajima T, Furukawa K, Ando T, Ishida H, Yoshida A, Nakamura Y, Kannagi R, Kiso M, Furukawa K: **Synthesis of disialyl Lewis a (Le(a)) structure in colon cancer cell lines by a sialyltransferase, ST6GalNAc VI, responsible for the synthesis of alpha-series gangliosides.** *J Biol Chem* 2003, **278(25):**22787-22794.
37. Dagan T, Sorek R, Sharon E, Ast G, Graur D: **AluGene: a database of Alu elements incorporated within protein-coding genes.** *Nucleic Acids Res* 2004:D489-D492.
38. Levy A, Sela N, Ast G: **TranspoGene and microTranspoGene: transposed elements influence on the transcriptome of seven vertebrates and invertebrates.** *Nucleic Acid Res* 2007 in press. [Epub ahead of print]